\documentclass[10pt,a4paper,floatfix]{article}

\usepackage{graphics}
\newcommand{\bea}{\begin{eqnarray}}
\newcommand{\eea}{\end{eqnarray}}
\newcommand{\be}{\begin{equation}}
\newcommand{\ee}{\end{equation}}

\def\be{\begin{eqnarray}}
\def\ee{\end{eqnarray}}
\def\bd{\begin{displaymath}}
\def\ed{\end{displaymath}}

\def\ga{\gamma}

\def\ADNDT{{At. Data Nucl. Data Tables }}

\def\NP{Nucl. Phys. }
\def\PR{Phys. Rev. }
\def\PRL{Phys. Rev. Lett. }
\def\PL{Phys. Lett. }
\def\jpg{J. Phys. G: Nucl. Part. Phys. }

\begin{document}
\title{Low energy ($p,\ga$) reactions in Ni and Cu nuclei 
using microscopic optical model}
%{($p,\ga$) reaction in Ni and Cu nuclei}
\author{ G. Gangopadhyay\\
Department of Physics, University of Calcutta\\
92, Acharya Prafulla Chandra Road, Kolkata-700 009, India}
\date{}
\maketitle

%\begin{abstract}
\centerline{\bf Abstract}
Radiative capture reactions for low energy protons have been theoretically
studied for Ni and Cu isotopes using the microscopic 
optical model. The optical potential has been obtained in the folding model
using different microscopic interactions with the nuclear densities from 
Relativistic Mean Field calculations. The calculated  total 
cross sections as well as the cross sections for individually low lying levels
have been compared with measurements involving stable nuclear targets. Rates for the rapid proton
capture process
have been evaluated for astrophysically important reactions.

%\end{abstract}

%\clearpage
%\section{Introduction}
\vskip 0.3cm

Optical model potentials constructed utilizing microscopic densities
from standard nuclear models have proved to be very successful in describing 
low energy nuclear reactions. Elastic scattering calculations using such
potentials have been able to explain the observed cross 
sections even in nuclei far off the stability valley. Low energy projectiles 
probe only the outermost part of the target nuclei. Hence the nuclear skin 
plays a very important role in such reactions. Theoretical models can provide
a good description of the density profile and is capable of producing 
excellent estimates
of reaction cross sections. Alternatively, the ability of different 
models to reproduce the nuclear density profile may be compared from their 
ability to predict reaction cross sections.

Proton capture reactions  at low energy are important to understand the 
astrophysical $rp$ process. At energies below the Coulomb barrier the cross 
sections are small. However, as the Gamow window lies entirely below the 
barrier, estimation of the cross section below the barrier is of crucial 
importance. We note that capture may lead to the ground state, to the excited 
states or to the continuum of the compound nucleus. 

Relativistic Mean Field (RMF) approach is now a standard tool in low energy
nuclear structure. It has been able to explain different features 
of stable and exotic nuclei like ground state binding energy, deformation, 
radius, excited states, spin-orbit splitting, neutron halo, etc\cite{RMF1}. 
RMF is known to provide a good description of various 
features in $A=60$ mass region [See \cite{CaNi} and Refs. therein]. 
There are different variations of the Lagrangian density as
well as a number of different parametrizations. In the present work we have 
employed
three such densities, NL3\cite{NL3}, TM1\cite{TM1}, and FSU Gold\cite{prl}, to study 
($p,\gamma)$ reactions in stable Ni and Cu nuclei. 
The NL3 density contains, apart 
from the usual terms for a nucleon meson system, nonlinear terms involving 
self coupling of scalar-isoscalar meson. The TM1 density includes additional
terms describing self-coupling of the vector-isoscalar meson. The FSU Gold 
density includes coupling between the vector-isoscalar meson and the 
vector-isovector meson as well. 
We note that results of our cross section calculation for all the three 
Lagrangian densities 
are practically identical and present the results for FSU Gold only.

In the conventional RMF+BCS approach, the equations obtained are solved under 
the assumptions of classical meson fields, time reversal symmetry, no-sea 
contribution, etc. Pairing is introduced under the BCS approximation. 
Usually the resulting equations are solved in a harmonic oscillator basis
\cite{Gam}. However, since we need the densities in co-ordinate space, a solution
of the Dirac and Klein Gordon equations in co-ordinate space has been preferred.
This approach has earlier been used 
\cite{CBe,CaNi,Zr} to study neutron rich nuclei in different mass regions. 
We have found that the present method describes the properties of the nuclei 
with $Z=28$ equally well as the more involved Relativistic Hartree Bogoliubov 
approach.
In the second and the third columns of Table 1, we compare the results for the 
binding
energy 
values for the 
stable isotopes for FSU Gold. The valence neutron proton 
correlation correction has been taken care of following the prescription of Ref.\cite{becor}. Much more important from the point of the density profile are the 
next two columns where we compare
the measured charge radii ($r_{ch}$) with
theory. The latter values have been obtained from the point proton distribution
($r_p$) using the simple prescription
$ r_{ch}=(r_p^2+0.64)^{1/2}$,
all quantities given in $fm$. The results show that RMF can describe the 
ground state of these nuclei with sufficient accuracy.

\begin{table}[ht]
\caption{Experimental binding energies\cite{AW} and radii\cite{radii} 
compared with calculated values for the FSU Gold Lagrangian density. The 
$G_{norm}$ values used in different isotopes 
are also indicated in the last two columns. See text for details.}
\begin{tabular}{lcccccc}\hline
&\multicolumn{2}{c}{B.E.(MeV)}&\multicolumn{2}{c}{$r_{ch}(fm)$} & \multicolumn{2}{c}{$G_{norm}$}\\
& Exp.&Theo.&Exp.&Theo.&JLM & DDM3Y\\\hline
$^{58}$Ni & 506.46 & 508.83 &3.775&3.751  & 0.85 & 0.85\\
$^{60}$Ni & 526.84 & 527.50 &3.812&3.779  & 0.60 & 0.60\\
$^{61}$Ni & 534.66 & 535.05 &3.822&3.792  & 0.70 & 0.70\\
$^{62}$Ni& 545.26 & 544.71 &3.841&2.828   & 0.60 & 0.60\\
$^{64}$Ni& 561.76 & 561.97 &3.859&3.827  & 0.95 & 0.80\\
$^{63}$Cu& 551.38 & 551.17 &3.883&3.848  & 0.55 & 0.55\\
$^{65}$Cu& 569.21 & 569.43&3.902&3.866 & 0.95 & 0.95\\ 
\hline
\end{tabular}
\end{table}

The optical model potentials for the reactions are obtained using two effective 
interactions
derived from the nuclear matter calculation in the local density approximation,
{\em i.e.} by substituting the nuclear matter density
with the density distribution of the finite nucleus.  
Thus the microscopic nuclear potentials have been obtained by folding the
effective interactions with the
microscopic densities from the RMF calculation.
The Coulomb potentials have been similarly generated by folding the Coulomb
interaction with the microscopic proton densities. 
We have already used such potentials
to calculate life times for proton, alpha and 
cluster radioactivity\cite{alp} as well as elastic proton scattering\cite{CBe}
in different mass regions of the periodic table.

One of the interactions chosen in the present work is the interaction of 
Jeukenne, Lejeune, and Mahaux (JLM)\cite{JLM74}  in which further 
improvement is incorporated in terms of the finite range of the effective
interaction by including a Gaussian form factor. We have used the global 
parameters for the effective interaction and the respective default 
normalizations for the potential components from Refs. \cite{MOMCS} and 
\cite{MOM} with Gaussian range values of 
$t_{real}=1.25~fm$ and $t_{imag}=1.35~ fm$. 

We have also used the density dependent interaction DDM3Y\cite{ddm3y1,ddm3y2} 
in the present work.
This was obtained from a finite range energy independent M3Y interaction by 
adding a zero range energy dependent pseudopotential and introducing a density 
dependent factor. This interaction has been employed widely in the study of 
nucleon nucleus as well as nucleus nucleus scattering, calculation of proton 
radioactivity, etc. The density dependence has been chosen in the form 
$C(1-\beta\rho^{2/3})$\cite{ddm3y2}. The constants were obtained from nuclear 
matter calculation\cite{ddm3y3} as $C=2.07$ and $\beta=1.624$ $fm^2$. 
For scattering we 
have taken  real and the imaginary parts of the potential as 0.9 times and 0.1 
times the DDM3Y potential, respectively. 

The reaction
calculations have been
performed with the computer code TALYS 1.2\cite{talys} assuming spherical 
symmetry for the target nuclei. The DDM3Y interaction is not a standard part 
of TALYS but can easily be incorporated.
Since nuclear matter-nucleon potential does not include a spin-orbit term, the 
TALYS 1.2 code obtains the spin-orbit potential
from the Scheerbaum prescription\cite{SO} coupled with the 
phenomenological complex potential depths $\lambda_{vso}$ and $\lambda_{wso}$.
\be U^{so}_{n(p)}(r)=(\lambda_{vso} +i\lambda_{wso})
\frac{1}{r}\frac{d}{dr}(\frac{2}{3}\rho_{p(n)}+\frac{1}{3}\rho_{n(p)})\ee 
The depths are functions of energy, given by $\lambda_{vso}=130\exp(-0.013E)
+40$ and $\lambda_{wso}=-0.2(E-20)$, $E$ in MeV. 
This has been used in the calculations of  both the interactions.

The TALYS code has a number of features useful to study reactions. We have 
employed the full Hauser-Feshbach calculation with transmission coefficients 
averaged over total angular momentum values and with corrections due to width 
fluctuations. Hilaire's microscopic level density values included in the 
code has been used though we have confirmed that the 
results are not substantially modified if a different level density formulae
is assumed.
Up to twenty five discrete levels of the compound nucleus have been 
included in the calculation. 
The gamma
ray strength has been calculated in the Hartree-Fock-Bogoliubov model. However, we find that though the trends have been correctly reproduced in all the cases, 
the actual values of the cross sections are often overpredicted. Thus the 
gamma ray strength 
was visually normalized to match with the experimentally observed cross 
sections using the parameter $G_{norm}$ in the code though no fit was performed.  In the last two columns of Table 1 we 
tabulate the values of this parameter used for the different targets.
We should also mention that in the case of $^{60,61}$Ni, the experimental 
values from different measurements differ by a large amount and  
we have chosen the latest measurements to determine $G_{norm}$.

In Figure 1, we have compared our results with various experimental measurements
in Ni isotopes and have found reasonable agreement. 
in Figure 2, we present the 
results for stable Cu isotopes. As the astrophysically important Gamow
window lies in the region 1.1 to 3.3 MeV for these nuclei, we compare the 
results up to 3.5 MeV proton energy. As already mentioned, 
$G_{norm}$ is the only parameter that we have modified to normalize
the experimental data.
All the other parameters in the Lagrangian density and the interaction are 
standard ones and have not been changed.
The DDM3Y and the JLM interactions
 perform almost identically in almost all the nuclei. The former
sometimes appears to produce slightly better results, but in view of the 
large disagreement between different measurements, this conclusion remains very 
tentative. We see that our calculation can explain cross section values ranging
over three orders of magnitude and also beyond the neutron evaporation 
threshold. We note here that the default local  and global optical potentials
\cite{omp} in the TALYS package also can be used with suitable normalization of
gamma ray strength to produce comparable 
results 
for certain energy ranges. For example, with $G_{norm}=0.5$, the results for 
the low energy values for DDM3Y and results using the default potentials are 
nearly identical in $^{64}$Ni($p,\ga$) reaction but above the neutron 
evaporation threshold, the predictions
by the default potentials, using the same $G_{norm}$ value, are definitely 
poorer compared to those of the microscopic calculations.

The cross-sections corresponding to the different low lying levels of the 
compound nucleus has been measured in some of the above reactions. In Figure 3, 
we show the results for the ground state and the first two excited states in 
the $^{63,65}$Cu($p,\ga$)$^{64,66}$Zn reactions using the inputs of the 
TALYS 1.2 code and the corresponding experimental measurements. Similar 
agreements are also observed in Ni isotopes. 
The results are for JLM interaction only. The DDM3Y results are nearly 
identical. We may conclude the present method to be suitable to describe 
the proton capture cross section by stable Ni and Cu isotopes.

With the success of the present approach, we have employed it to calculate
the astrophysical rapid proton capture rate in Ni and Cu nuclei. 
Nucleosynthesis theories\cite{nuc} suggest that the above process is very important
in $^{56}$Ni and $^{57}$Cu for which we present 
our results in Figure 4. Since the laboratory  cross sections are not 
available for the two unstable targets, we have assumed $G_{norm}=1$. We also 
compare our results with two 
theoretical calculations, based on the Hauser-Feshbach formalism code 
NON-SMOKER\cite{Smoker} and Shell model\cite{SM}, respectively. The stellar enhancement 
factor has not been incorporated in the results. The results for DDM3Y 
interaction are nearly identical and have not been plotted. We note that there 
are substantial differences between the three calculations in the case of $^{58}$Ni, 
particularly the NON-SMOKER results being much larger compared to the present 
ones. We find that the cross sections from the NON-SMOKER code
\cite{Smoker} are very much larger than experimental measurements 
as one goes to proton rich Ni isotopes. Thus we may expect the astrophysical 
rates from \cite{Smoker} to be greater in $^{56}$Ni.

In summary, cross sections for low energy ($p,\ga$) reactions for stable 
Ni and Cu nuclei have been 
calculated using the TALYS code. The microscopic optical potential has been 
obtained by folding two different 
microscopic interactions, JLM and DDM3Y, with the densities of the target
nuclei obtained from three different RMF Lagrangian densities, {\em viz.} 
NL3, TM1, and FSU Gold. 
Astrophysical rates for the $rp$ process  have been calculates and compared with
standard calculations in two important nuclei $^{56}$Ni and $^{57}$Cu.

\section*{Acknowledgement}

This work has been carried out with financial assistance of the UGC sponsored
DRS Programme of the Department of Physics of the University of Calcutta.
The author gratefully acknowledges the hospitality of the ICTP, Trieste where a 
part of the work was carried out.

\begin{figure}[th]
\parindent -2cm
\vskip -4cm
\resizebox{13cm}{!}{\includegraphics{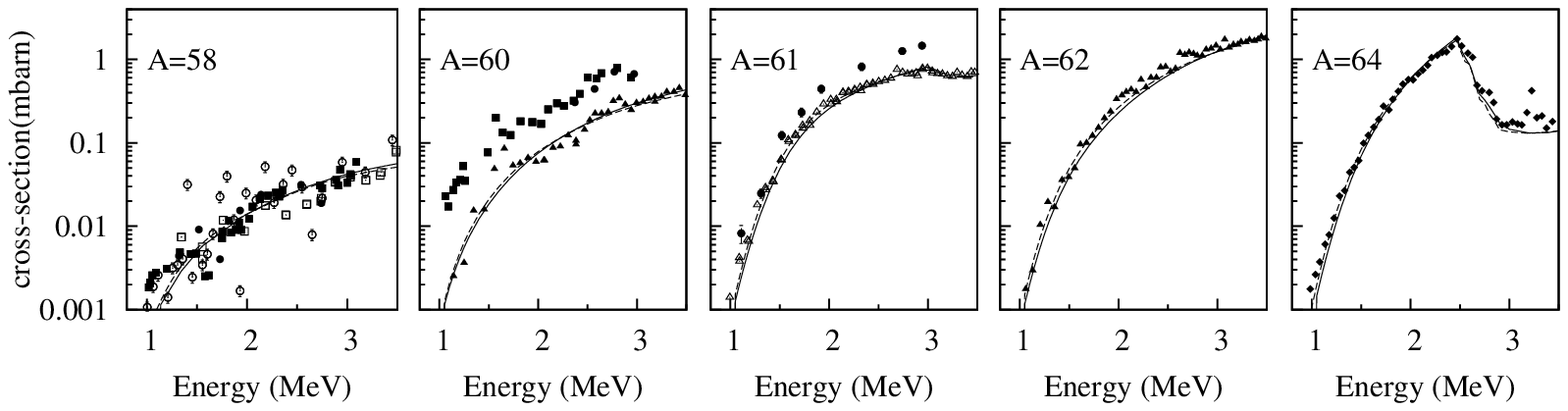}}
\caption{Cross sections for ($p,\ga$) reactions in stable Ni isotopes. The 
mass numbers of the target nuclei are indicated.  The data are form Refs. 
\cite{ni58p} (open square), \cite{ni58p1} (filled square),  \cite{ni58p3} 
(filled circle), \cite{ni58p4} (open circle),  \cite{ni60p2} (filled triangle),
\cite{ni61p} (open triangle) and
\cite{ni64p} (diamond). The solid and the dashed lines refer 
to results for JLM and DDM3Y interactions, respectively.}
%\end{figure}
%\begin{figure}[bh]
\center

\vskip -4cm
\resizebox{!}{!}{\includegraphics{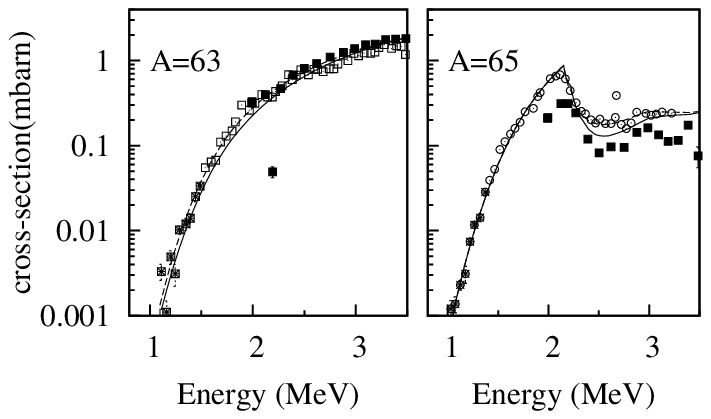}}
\caption{Cross sections for ($p,\ga$) reactions in stable Cu isotopes. 
The data are from \cite{ni64p}(open square),\cite{cu63p3}(filled square) and 
\cite{ni64pexst} (open circle). See caption of Figure 1 for details.} 
\end{figure}
\begin{figure}[t]
\vskip -4cm
\center
\resizebox{!}{!}{\includegraphics{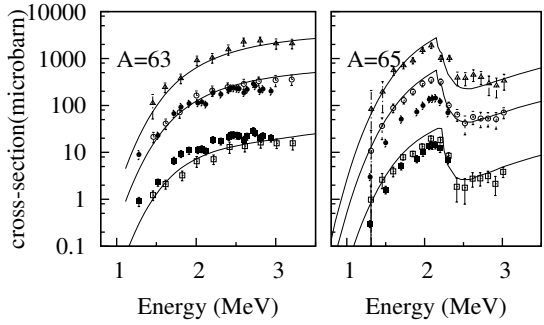}}
\caption{Partial cross sections for $^{63,65}$Cu($p,\ga$) reactions to the low 
lying  
states in $^{64,66}$Zn. Open (filled) symbols 
refer to data from \cite{cu63p4}(\cite{cu63pex1}). Squares, circles and 
triangles represent data for transition to the ground state and to the first 
excited state (multiplied by 10) and the second excited state (multiplied by 
100), respectively. The mass numbers of the target nuclei are indicated.}
\end{figure}
\begin{figure}[b]

\vskip -4cm
\resizebox{13cm}{!}{\includegraphics{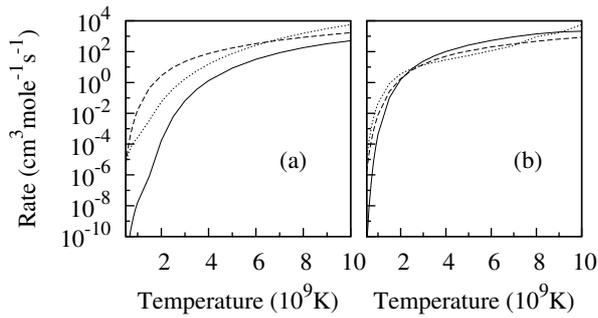}}
\caption{Astrophysical proton capture rates in (a) $^{56}$Ni and (b)$^{57}$Cu
 given by present work (solid line), NON-SMOKER calculation
\cite{Smoker}(dashed line) and Shell Model results\cite{SM}
(dotted line).}
\end{figure}
\end{document}